# Design of Phosphorene for Hydrogen Evolution Performance Comparable to Platinum

Yongqing Cai[§], Junfeng Gao[‡], Shuai Chen[†], Qingqing Ke[¶], Gang Zhang[†],[*], and Yong-Wei Zhang[†][*]

[§]Joint Key Laboratory of the Ministry of Education, Institute of Applied Physics and Materials Engineering, University of Macau, Taipa, Macau, China

[‡]Key Laboratory of Materials Modification by Laser, Ion and Electron Beams, Dalian University of Technology, Dalian, China 116024

[†]Institute of High Performance Computing, A*STAR, Singapore 138632

[¶]Institute of Materials Research and Engineering, A*STAR, Singapore 138634

**ABSTRACT:** Phosphorene, a monolayer of bulk black phosphorus, is promising for light harvest owing to its high charge mobility and tunable direct band gap covering a broad spectral range of light. Here, via atomic-scale first-principles simulations, we report a ultrahigh activity of hydrogen evolution reaction (HER) of phosphorene originated from defective activation. Quantitative evaluation of the Gibbs free energy of the ad/desorption of hydrogen (H*) to/from phosphorene ($\Delta G_{H^*}$) reveals that atomic vacancies and edges play a dominant role in activating the reaction. We find that the defective states, empty and well-localized around the defect core, are compensated by H* species. This induces a proper hydrogen interaction complying with the thermoneutral condition of the free energy ($\Delta G_{H^*} \approx 0$) comparable to platinum. Our findings of the highly activating defective states suggest the design of non-metal HER catalysts with structural engineering of earth-abundant phosphorus structures.

**Introduction**

Hydrogen, which is the cleanest fuel, represents one of the most promising energy sources that potentially reduce the fossil fuel dependence. Nowadays, 90% of hydrogen is produced through reformation of fossil fuels, which is unsustainable and accompanied by severe $CO_2$ emissions.[1] Thus, exploring alternative routes to efficiently, cost-effectively and cleanly produce hydrogen is an active research topic. Hydrogen evolution reaction (HER, 2 $H^+$ + 2 $e^-$ → $H_2$), which is the cathodic reaction in electrochemical water splitting, offers the potential to produce $H_2$ in such a manner.[2] State-of-the-art HER catalysts mainly consist of noble metals, in particular, Pt, owing to their highly efficient functionality.[3] However, the scarcity and high cost of noble metals limit their scalable applications and prompt the searching for low-cost alternatives.

Considerable progress has been made in recent years in exploring two-dimensional (2D) materials, especially transition metal dichalcogenides (TMDs) like $MoS_2$.[4,5] More recently, metal-free catalysts consisting of solely elemental phosphorus have shown a great potential in hydrogen production.[6,7] Phosphorus is one of the most abundant elements on the earth (up to 100 billion tons[8]) and presents a great flexibility in crystalline structures (more than six polymorphs at ambient condition).[9] For example, fibrous red phosphorus, an elemental semiconductor, was reported to have a record high HER rate (684 $\mu mol h^{-1} g^{-1}$), and a high stability in water under light irradiation.[6] Currently, there is a growing interest in exploring other phosphorus polymorphs for high HER performance.

Black phosphorus, the most stable allotrope of the element phosphorus with a layered structure, is a fascinating material that is promising for various applications in optoelectronics,[10,11] gas sensing,[12] and more recently electrochemical catalysis.[13-15] With a thickness-dependent direct band gap from 1.52 eV for monolayer (known as phosphorene) to 0.39 eV for bulk,[16] and a high carrier transport efficiency, phosphorene is an ideal candidate as a photocatalyst for broadband solar absorption.[17,18] It has a poor stability upon exposure to air, which was previously attributed to the presence of water and oxygen. However, recent experimental studies showed that phosphorene could be exfoliated in liquid solvents[19,20] and stable in deoxygenated water.[21,22] Considering the densely packed lone-pair $p$ electrons of phosphorus atoms in the phosphorene sheet,[23] it was thought that the hydrogen would have an unfavorable adsorption above its electron-rich surfaces related to lone-pair states of phosphorus. Recently, through forming nanohybrid with graphitic carbon nitride or CdS, phosphorene co-catalyst was found to have a good HER activity as a metal-free photocatalyst from visible to near-infrared range.[24,25] To our knowledge, the mechanism underlying this fascinating HER process remains unclear.[26] As the understandings for HER in 2D materials so far were mainly built for systems containing transition metals containing $d$ orbitals, it is expected that the mechanism underlying the HER activity in phosphorene, which is a typical $s$-$p$ hybridized system, should be different from that of transition metals. Clearly, understanding how the hydrogen atoms interact with the phosphorus atoms in phosphorene to produce $H_2$ is highly desired.

In this work, based on a thermodynamic descriptor-based approach and using density functional theory (DFT), we show that while perfect phosphorene sheet is catalytically inert, phosphorene with atomic defects is able to serve as a superior HER catalyst. Via a thorough screening of atomic sites in vacancy- and/or edge-contained phosphorene, the active sites for HER process are identified. Through examining the Gibbs free energy, which is a key metric for evaluating the HER activity, we reveal that the odd-sized vacancies like monovacancy and trivacancy contain several active sites that show a superior catalytic ability, comparable to that of platinum. Although atomic sites at divacancy are intrinsically inert, a proper tensile strain can make it catalytically active. We further reveal that the superior HER activity in phosphorene arises from the localized states of strongly distorted bonds at those defects. Both the charge transfer from the hydrogen atoms to the defects and the locally distorted lattice sites in the defect cores account for the favorable hydrogen-phosphorene interaction and the superior HER.

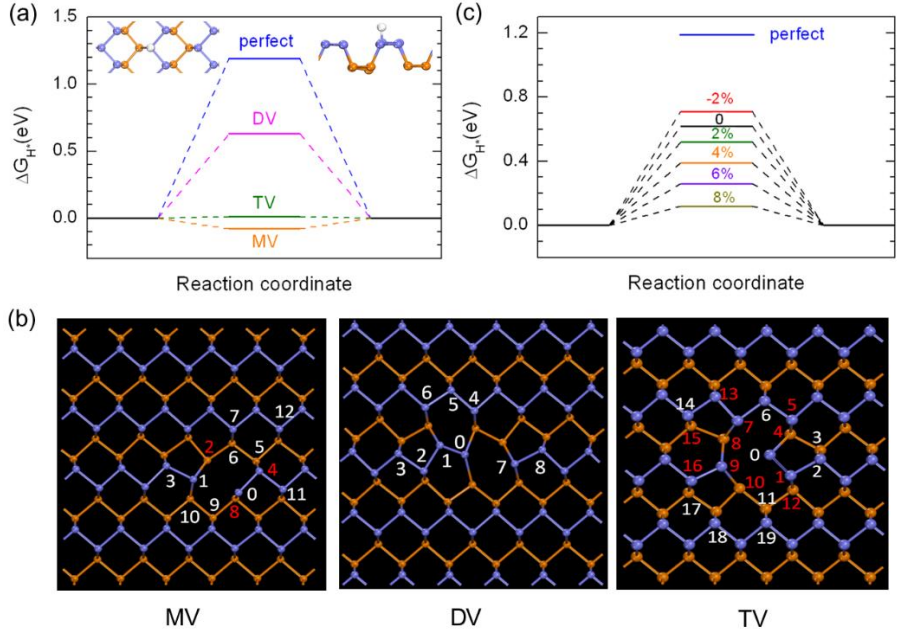

**Figure 1.** High HER activity associated with atomic vacancies in phosphorene. (a) Comparison of the free energy diagram in the most favored sites for perfect, MV, DV, and TV cases. Inset shows the geometries of hydrogen adsorbed above pristine phosphorene. (b) Atomic models for the MV, DV, and TV in phosphorene. The red (white) numbers beside atoms indicate a relatively high (low) HER activity at this site. (c) Strain modulation of the free energy for H adsorption on atom 0 of the DV core as shown in (b).

**Computational methodology**

**First-principles structural optimization and energetics calculations:** We perform spin-polarized first-principles calculations by using Vienna ab initio simulation package (VASP)[27] with the Perdew–Burke–Ernzerhof (PBE) functional being selected. Screening of the adsorbing states of hydrogen on perfect and phosphorus-deficient phosphorene is modeled by creating a 5×4 supercell of monolayer phosphorene with relaxed lattice constants of 3.305 and 4.617 Å by DFT, consistent with previous study.[10] A vacuum layer of thickness greater than 15 Å is adopted. A cutoff of 400 eV for treating the kinetic energy and electronic density and a 3×3×1 (1×4×1) Monkhorst-Pack grid are used for point (edge) defect. All the structures are fully relaxed until the force on each atom is less than 0.005 eV/Å. The charge transfer analysis is performed by calculating the differential charge density $\Delta\rho(\mathbf{r})$, which is defined as $\Delta\rho(\mathbf{r})=\rho_{\text{phosphorene+H}}(\mathbf{r})-\rho_{\text{phosphorene}}(\mathbf{r})-\rho_{\text{H}}(\mathbf{r})$, where $\rho_{\text{phosphorene+H}}(\mathbf{r})$, $\rho_{\text{phosphorene}}(\mathbf{r})$, and $\rho_{\text{H}}(\mathbf{r})$ are the charge densities of the H adsorbed phosphorene, phosphorene, and H atom, respectively.

**Details of Gibbs free energy calculation for HER**: According to the definition of Gibbs free energy $\Delta G_{H^*}=\Delta E_{H^*}+\Delta E_{ZPE}-T\Delta S_{H^*}$ where $\Delta E_{ZPE}$ is the correction of zero point energy, and $\Delta S_{H^*}$ is the entropy difference between the adsorbed state and the gas state, and $T$ is temperature set to 300 K, and $\Delta E_{H^*}$ is the H chemisorption energy, which is defined as $\Delta E_{H^*} = \frac{1}{n}(E_{nH^*\text{-Cat}} - E_{\text{Cat}} - \frac{n}{2}E_{H_2})$, where $n$ is the number of H atoms in the supercell. The calculation of $\Delta G_{H^*}$ thus can be decomposed into the individual calculations of $\Delta E_{H^*}$, $\Delta E_{ZPE}$, and correction of the vibrational entropy of the adsorption $\Delta S_{H^*}$. The $\Delta E_{H^*}$ is obtained from the prediction of the total energies of the H adsorbed catalysis ($E_{nH^*\text{-Cat}}$), pristine catalysis ($E_{\text{Cat}}$), and isolated $H_2$ molecule ($E_{H_2}$) by the first-principles calculation. The zero point energy $\Delta E_{ZPE}$ requires the calculation of the frequency of the free $H_2$ molecule, and that of the adsorbed $H^*$. For instance, in the case of phosphorene, the vibrational frequency of $H^*$ has three eigenvalues: 2307.95, 746.37, and 608.51 cm$^{-1}$ through solving the real space Hessian matrix. As the vibrational frequency of free $H_2$ is 4401.21 cm$^{-1}$, the $\Delta E_{ZPE}$ is computed as $\Delta E_{ZPE}= \sum \hbar\omega_i(H^*)-\hbar\omega(H_2)/2=0.08$ eV, where ℏ is the reduced Planck constant. In the last term of the Gibbs free energy, the $\Delta S_{H^*}$ term is the difference between the vibrational entropy of the adsorbed $H^*$ and that of the gas $H_2$ phase. Since the configuration entropy of the solid $H^*$ is small and can be neglected, the $\Delta S_{H^*}$ can be approximated by $\Delta S_{H^*} \approx -S_{H_2}/2$, where $S_{H_2}$ is the entropy of the gas molecule hydrogen. According to Ref. 28, under the standard condition (1 bar of $H_2$, pH=0, and T=300 K), $S_{H_2}$ is $1.35\times10^{-3}$ eV/K, and thus $T\Delta S_{H^*}$ is -0.21 eV.

**Results and discussion**

For HER in acid solution on a catalyst (Cat), the first step is the adsorption of the hydrogen to the catalyst (Volmer process), which is described as $H^+ + e^- + \text{Cat} \leftrightarrow H^*\text{-Cat}$, where the $H^*$ denotes an adsorbed hydrogen (H) atom. The second step is the formation and release of the hydrogen molecule via H*-Cat + H*-Cat ↔ H$_2$ + 2Cat (Tafel process) or H$^+$ + e$^-$ + H*-Cat ↔ H$_2$ + Cat (Heyrovsky process). According to Nørskov *et. al.*,[29] the above processes for the hydrogen transformation can be thermodynamically described by the Gibbs free energy ($\Delta G_{H^*}$). A good catalyst for HER should bind the H neither too weakly nor too strongly.[30,31] A facile replenishing of catalyst with H means a reversible adsorption/desorption of H and $\Delta G_{H^*}$ should comply with the thermoneutral condition ($\Delta G_{H^*} \approx 0$).[29]

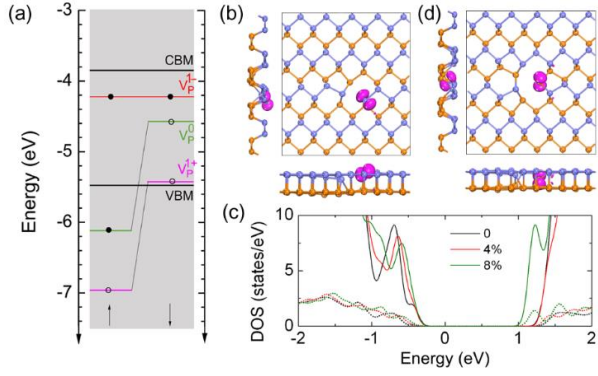

**Figure 2**. Spatial and energetic distributions of defective states in phosphorene obtained by HSE method to accurately account for the electronic self-interaction and the band gap. (a) Alignment of defective levels of MV in different charged states (+1, 0, -1) relative to the band edges of perfect phosphorene. (b) Different views of the spatial distribution of the unoccupied defective state (~0.2 eV above VBM) of the neutral MV. (c) Evolution of the total DOS (solid lines) and the local DOS for the defective core atoms (dotted lines) of the DV-contained system with strain. (d) Spatial distribution of the unoccupied defective state (~0.2 eV above VBM) of the neutral TV.

**Table 1. Free energy (eV) for H adsorption at the atomic sites (integer number) of vacancies as labeled in Figure 1.**

|  | Single H adsorption | | | Double H adsorption | |
|---|---|---|---|---|---|
| Sites | MV | DV | TV | Sites | |
| 0 | -0.47 | 0.69 | -0.43 | MV | |
| 1 | -0.40 | 1.12 | **0.21** | 0-2 | **0.05** |
| 2 | **-0.02** | 1.41 | 0.32 | 0-4 | 0.47 |
| 3 | -0.41 | 1.45 | 0.32 | 0-8 | 0.33 |
| 4 | **0.17** | 0.78 | **0.21** | DV | |
| 5 | 0.58 | 0.92 | **0.18** | 0-1 | 0.55 |
| 6 | 0.50 | 1.04 | 0.27 | 0-4 | 0.57 |
| 7 | 0.50 | 0.99 | **0.07** | 0-7 | 0.81 |
| 8 | **0.14** | 1.17 | **0.11** | TV | |
| 9 | 0.44 |  | **0.11** | 0-7 | **0.01** |
| 10 | 0.38 |  | **0.07** | 0-8 | **-0.10** |
| 11 | 0.41 |  | 0.27 | 0-9 | **-0.07** |
| 12 | 0.67 |  | **0.18** | 0-10 | **-0.01** |

By using DFT calculations, we first examine the adsorption on a perfect monolayer phosphorene under dilute coverage limit (a surface coverage of 0.025). The most stable configuration adopts a nearly vertical geometry with the H-P bond being aligned nearly normal to the basal plane. The $\Delta G_{H^*}$ (Figure 1a) is found to be 1.25 eV and thus the process is endothermic. When increasing the coverage by adsorbing two H atoms simultaneously, the $\Delta G_{H^*}$ decreases slightly to around 0.96 eV (0.99 eV) with the two H atoms distributed along the armchair (zigzag) direction. Therefore, perfect phosphorene is essentially inactive due to the large thermodynamically uphill step of the adsorption.

It is well-known that the presence of atomic defects like edges and vacancies is able to promote HER activity of TMDs.[32] Herein, we systemically examine the HER activity associated with the three most popular vacancy defects in phosphorene: monovacancy (MV), divacancy (DV), and trivacancy (TV). Figure 1b shows the lowest energy configurations of the MV and DV with pentagon-nonagon (59) pairs and double pentagon-heptagon (5757-A) pairs, respectively.[33] We have fully sampled all the core atoms (labeled and marked with numbers in Figure 1b) by considering the local site symmetry of each defect, ranging from atom 0 to 12 for the MV case and atom 0 to 8 for the DV case (note the inversion symmetry in the 5757-A type DV). The values of the $\Delta G_{H^*}$ are compiled in Table 1. Clearly introduction of the MV overall increases the binding, which is signified by a dramatic decrease of $\Delta G_{H^*}$. It is not surprising that the strongest adsorption occurs at the dangling atom (atom 0) due to the saturation of the breaking bonds. However, the adsorption at this site is too strong ($\Delta G_{H^*}$ of -0.47 eV) for HER.

Interestingly, the nearest neighbors of the dangling atom 0 show high activity with $\Delta G_{H^*}$ of -0.02, 0.17, and 0.14 eV for sites 2, 4, and 8, respectively. For other atomic sites around the core of MV, the activity is poor as the binding energy is either too strong, such as sites 1 and 3 with $\Delta G_{H^*}$ around -0.40 eV, or too weak, such as sites 5-7, and 9-12 with $\Delta G_{H^*}$ around 0.50 eV (Table 1). In real situation, hydrogen may preferably adsorb at the dangling 0 site which may change the HER activity at other sites. To examine this issue, we also consider concurrent adsorption of two hydrogens at 0 and 2 sites, 0 and 4 sites, and 0 and 8 sites (see Table 1). Indeed, the $\Delta G_{H^*}$ for latter two combinations are 0.47 and 0.33 eV, respectively. However, the simultaneous adsorption at 0 and 2 sites still comply with the thermoneutral condition with $\Delta G_{H^*}$ of 0.05 eV.

In reality, two isolated MVs may coalesce to form a DV, which may show a different HER activity in comparison with MV due to the structural difference. Indeed, we find that the atoms around the core of a DV are all inactive. The values of $\Delta G_{H^*}$ range from 0.69 (0 site) to 1.45 eV (3 site) (see Table 1), indicating an endothermic adsorption, which is the same as perfect phosphorene. The trend is also true for increasing the uptake of two hydrogen atoms. The relatively weak interaction of DV with H atom can be ascribed to a fully passivated structure with all atoms having a coordination number of three, similar to that of perfect phosphorene. Nevertheless, by applying equal biaxial strains from -2% (compressive) to +8% (tensile), the HER activity of DV is found to be highly tunable. We have only considered moderate compressive strain up to -2% as the 2D phosphorene sheet tends to form ripples at higher strains. The $\Delta G_{H^*}$ decreases gradually with strain for atoms 0, 1, 4-6, while fluctuating for atoms 2, 3 (see details in Table S1 in Supporting Information). Figure 1c depicts the change of $\Delta G_{H^*}$ with strain for adsorption at atom 0, where the $\Delta G_{H^*}$ drops steadily from 0.69 eV at zero strain to 0.14 eV at +8% strain. A similar enhanced activity is also found for adsorption at atom 4: the $\Delta G_{H^*}$ decreases to 0.07 eV at +8% strain. Our findings indicate that strain engineering is effective in enhancing the HER of DV-contained phosphorene. Moreover, the +8% strain needed for good HER activity is far below the failure strain limit of phosphorene (30% from Ref. 34) and can be realized in the concave region of rippled phosphorene.[35] Surprisingly, the HER activities of perfect phosphorene are found to be insensitive to the strain engineering and those of MV under strain change only slightly (see details in Tables S2, S3 in Supporting Information).

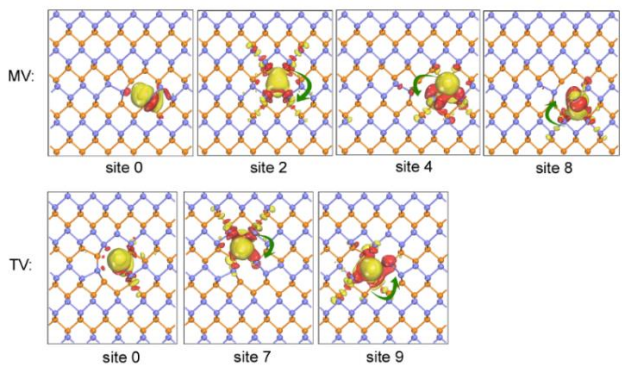

**Figure 3**. Isosurface plot of differential charge density for H atom adsorbed at several sites of MV (sites 0, 2, 4 and 8) and TV (sites 0, 7 and 9) of phosphorene. The red (yellow) color represents the accumulation (depletion) of electrons. The green arrows indicate the direction of charge transfer from the H atom to the unsaturated P atom in the core centers of the MV and TV.

It is known that in TMDs, the $d$ orbitals of the exposed transitional cations play a critical role in the vacancy-promoted HER activation.[32] This scenario is absent here due to the absence of the $d$ states in phosphorene. Instead, we show that there is a different mechanism involved. To explain the activating mechanism associated with the MV defect, we investigate the local electronic structure of the MV. Interestingly, the band structure and density of states (DOS) of the MV-contained structure show a spin-split and singly occupied defective level (denoted as $a_1$) (see Figure S1 in Supporting Information). Figure 2a shows the energy positions of the $a_1$ level of MV (its real space projection is shown in Figure 2b) in different charged states, together with the valence band maximum (VBM) and conduction band minimum (CBM) of perfect phosphorene calculated via HSE hybrid functional method. To facilitate the comparison of the alignment of the +1($V_P^{+1}$), 0($V_P^0$), -1($V_P^{-1}$) charged states of MV, all the levels are referenced to the vacuum potential. Based on the fully relaxed structure by HSE calculation, the MV is stable in the -1 state, in which the empty (occupied) $a_1$ level is ~1.2 eV above the VBM (~0.4 eV below the CBM). In contrast, the spin-up component of $a_1$ level for the $V_P^{+1}$ state is empty and located below the VBM, indicating that +1 charged state is unstable since the empty level cannot lie below the Fermi level. However, the $V_P^{+1}$ state would be possible if one or two holes being bound with the $V_P^0$ or $V_P^{-1}$ via Coulomb attraction and forming a series of hydrogenic states above the VBM. Previous study showed that the MV is more energetically favorable in -1 state[36] and serves as the p-type conduction.[37]

For the case of DV, there is no dangling state in the band gap (Figure 2c). However, as substantial strain fields pre-exist at the core atoms due to the bond reconstruction,[33] further stretching would weaken these atomic bonds. This triggers a continuous narrowing of the gap of the bonding and antibonding states of the core atoms with tensile strain (see dotted lines for 0, +4% and 8% strains in Figure 2c). The charge becomes gradually localized at the nuclei with the increased lattice spacing, which is the underlying reason for the enhanced interaction and the promoted HER activity.

Besides the MV and DV, we also examine the TV defect. Similar to the MV case, the TV structure created by a loss of odd (three) phosphorus atoms cannot be fully passivated through bond reconstruction. The most stable TV structure is shown in Figure 1b, where the dangling atom is numbered as 0, and appears similar defective state (Figure 2d) as MV. We perform a thorough study of the core atoms (atoms 0-19), and the corresponding values of $\Delta G_{H*}$ are compiled in Table 1 for atoms 0-12 and the rest are given in Table S4 in Supporting Information. The strongest binding occurs at the dangling atom 0 with the $\Delta G_{H*}$ of -0.47 eV, which is unfavorable for the release of hydrogen. However, in its direct neighbors (atoms 7-10), a high activity with $\Delta G_{H*}$ between 0.07 and 0.11 eV is found, which nearly comply with the thermoneutral criterion. Importantly, such high activity at these sites remains when a hydrogen atom is pre-adsorbed at the dangling atom (0 site) (Table 1 and Table S5 in Supporting Information). Notably, arrangement of atoms 6-11 around the big hole of the vacancy resembles an edge-like structure.

Through calculating the differential charge density, we find that electrons are transferred from the H atom to the dangling atom (site 0 in Figure 3). The atomic sites 0, 2, 4, and 8 are selected to represent the cases of strong binding (site 0) and weak interactions (sites 2, 4, and 8). It is seen that for all the four sites, the H atom donates electrons to the respective connected P atoms. Interestingly, for the weak binding at sites 2, 4, 8, a clear charge transfer from the H atom to the dangling atom 0 is observed although the H atom is not bonded to atom 0 (see the green arrow in Figure 3). Therefore, the H atom acts as an effective donor, which can effectively compensate the defect and induce a negatively charged MV. Compared with perfect phosphorene, this interaction decreases the uphill energy in the $\Delta G_{H*}$ diagram for HER at MV. Another reason for the high activity of these sites could be due to the local lattice distortions. Previous works revealed that atomic vacancies in phosphorene induce a strain field and bond deformation like elongation around the defect core.[33,38] The tensile strain is known to weaken the atomic bonding in phosphorene via upward (downward) shift of bonding (antibonding) orbitals.[39] This could enhance the hydrogen-phosphorus interaction via hybridization of the s-orbital of H and the anchored phosphorus atoms, leading to an appropriate coupling that favors HER and accounts for the promoted interaction of dopants and the sheet.[38] Similarly, mechanism underlying the high HER activity for the TV case is the same as the aforementioned MV case with a partial compensation of the localized dangling states (Figure 2d) due to the adsorption of the H atom, which induces a charge transfer to the phosphorus atom (see Figure 3).

Motivated by the high activity of the atoms at the edge-liked part of the TV structure as discussed in the preceding section, we next examine the role of phosphorene edges in HER. In the case of TMDs, the edges largely govern their HER performance.[40-42] Figure 4a shows the relaxed armchair phosphorene nanoribbon. To identify the edge effect on H adsorption, we have considered the change of energetics with the binding sites, which are represented by integer $R_N$, sequential numbering of the atomic sites along the ribbon width direction. The corresponding values of $\Delta G_{H*}$ are plotted in Figure 4b. Consistent with the pristine phosphorene results (Figure 1), the $\Delta G_{H*}$ is 1.20 eV for single H adsorption at the middle of the ribbon ($R_N$=9). With decreasing $R_N$, the $\Delta G_{H*}$ decreases gradually and reaches -0.22 eV for $R_N$=1 at the edge, which is a little overbound, but still good for HER. The continuous drop of the $\Delta G_{H*}$ towards the edge is indicative of a diminishing role of the endothermic adsorption at the interior bulk.

We calculate the band structure and DOS of armchair-edge terminated nanoribbon of phosphorene (Figure 5a and b). Our calculation reveals plenty of empty edge state (EES) and occupied edge state (OES). While the OES distributes largely in the interior of the ribbon, the EES is highly localized at the unpassivated edge atoms. The EES highly resembles the a1 defective state of MV from both energetic and spatial perspectives. The favorable H adsorption and good HER activity at edge can be attributed to the presence of p-orbital- dominated EES. Charge transfer from H to the shallow EES (see Figure 5c) induces a bound donor-defect complex. We also consider two high coverages of H, that is, 75% and 100% (Figure 4c and d) of exposed edge atoms of phosphorene with concurrent uptake of H atoms, and the $\Delta G_{H^*}$ is 0.02 and -0.07 eV, respectively, nearly complying with the thermoneutral condition.

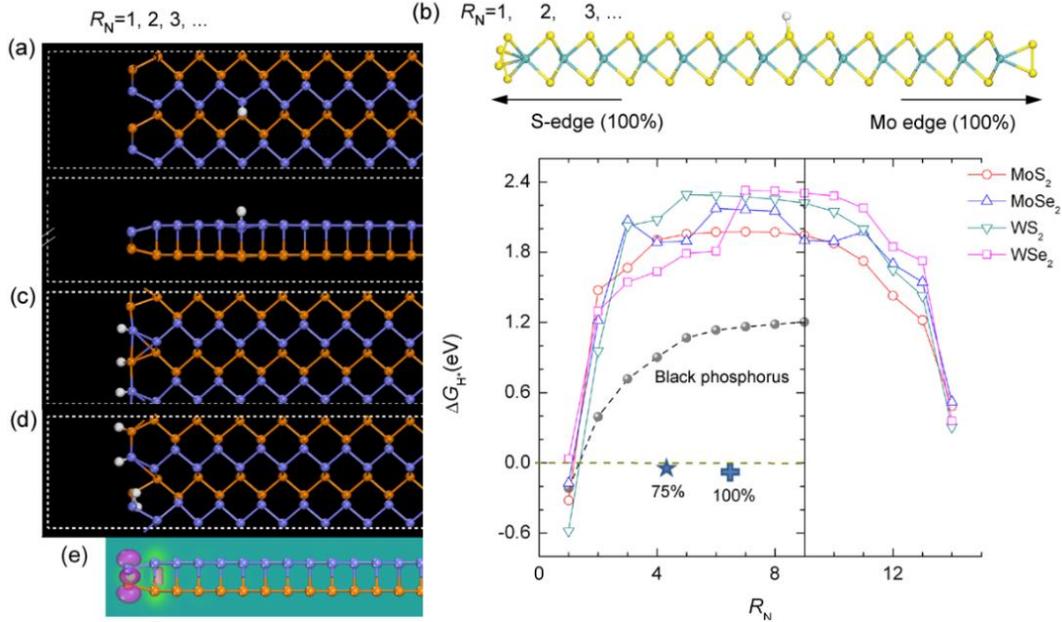

**Figure 4.** Effect of phosphorene edge on the HER activity. (a) Atomic model for the site-dependent adsorption of H at an atomic state from armchair edge to the interior sheet. (b) Comparison of HER between phosphorene edge and the semiconducting TMD (MoS$_2$, MoSe$_2$, WS$_2$, WSe$_2$) edges. The top inset corresponds to the atomic models (zigzag nanoribbon) used for simulating the S- and Mo- edge of the MoS$_2$ and the subsequent H sampling across the nanoribbon. The star marks represent concurrent uptake of H atoms with 75% and 100% of exposed edge phosphorus atoms of phosphorene adsorbed with H ($R_N$=1), and the atomic models are shown in (c) and (d), respectively. (e) Side view of the edge states in phosphorene.

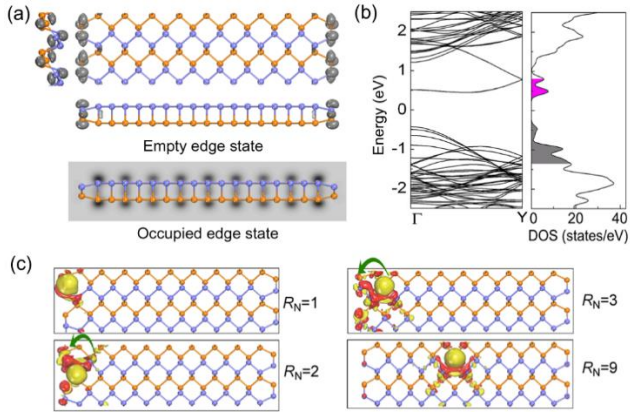

**Figure 5.** (a) Distribution of the empty edge state (EES) and occupied edge state (OES) of the phosphorene armchair nanoribbon. (b) Band structure and DOS for the phosphorene armchair nanoribbon. The shaded part in red (grey) color corresponds to the EES (OES). (c) Charge transfer between hydrogen and armchair edge (see the definition of $R_N$ in Figure 4) of phosphorene. Noted that all the results are derived by PBE calculation.

Our calculation indicates that pre-absorption of some H atoms or at a high coverage of H atom at edge can effectively weaken the interaction between hydrogen and edge. This is understandable since a saturation of the unpassivated edge with H is able to suppress the localized edge states. Moreover, with a high H coverage, the repulsive electrostatic interaction between the adsorbed H atoms can exert an energy penalty, which outweighs the energy decrease arising from orbital hybridizations.

To highlight the unique *p*-orbital-driven HER performance in phosphorene edge, we also investigate the edge effect of TMDs materials (MoS$_2$, MoSe$_2$, WS$_2$, and WSe$_2$), which are well-known for their *d*-orbital domination at band edges. Since the zigzag edge of TMDs is generally more energetically favorable and frequently observed experimentally[43], we have thus chosen the zigzag edge in the study. As a representative of the four MX$_2$ materials (M=Mo/W, X=S/Se), the zigzag edge of MoS$_2$ is plotted in Figure 4b (top panel). Due to the lack of inversion symmetry, the zigzag edge of MX$_2$ has two different terminations: the X-edge and M-edge.[43] Both edges possess several derivatives with different contents of edge exposed sulfur atoms.[43] Herein we choose a typical zigzag ribbon with a width of 15 X-M dimmer lines, together with a 1×2 supercell along the periodic direction and a single H adsorption. The two edges of the ribbon correspond to X-edge and M-edge, respectively. We find that within the interior part of the sheet (for H adsorption at $R_N$=9), all the TMDs show an inactive performance with the $\Delta G_{H^*}$ of 1.94, 1.90, 2.22, and 2.31 eV for MoS$_2$, MoSe$_2$, WS$_2$ and WSe$_2$, respectively, which are much larger than the value of 1.20 eV for phosphorene. Note that our result for MoS$_2$ is consistent with previous calculations of ~2.0 eV.[32] For H adsorption at the X-edge (M-edge), the $\Delta G_{H^*}$ changes significantly to -0.32 eV (0.48 eV), -0.17 eV (0.52 eV), -0.58 eV (0.31 eV) and 0.03 eV (0.36 eV) for MoS$_2$, MoSe$_2$, WS$_2$ and WSe$_2$, respectively. It can be seen that the edged phosphorene shows a comparable or even better performance compared to the TMD materials (see detailed values in Table S6 in Supporting Information), partially due to the remarkable edge state as shown in Figure 4e. As shown in Figure 4b, there exists a flatter trend of site-dependent $\Delta G_{H*}$ at the edge of phosphorene ribbon than that of the TMDs, indicating a greater potential for the edge atoms of phosphorene for HER.

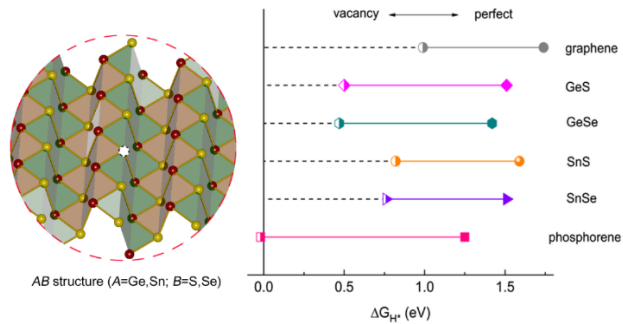

**Figure 6**. Comparison of the HER activity of phosphorene with graphene and other phosphorene-like *AB* (*A*=Ge, Sn; *B*=S, Se) materials.

We have revealed that defects such as vacancies and edges are able to induce a superior HER activity in phosphorene, and the underlying reasons can be attributed to the favored H-P hybridization due to local bond distortions around the defects and a partial compensation of the localized defective states of vacancies and edges by the hydrogen species. It is natural to ask whether this defect-promoted superior HER is a general or rare behavior of 2D materials. To facilitate a direct comparison with phosphorene, we choose graphene, a predecessor of phosphorene, and several phosphorene-like metal monochalcogenides AB (A=Ge, Sn, B=S, Se) 2D sheets (Figure 6), which have a similar puckered structure as phosphorene. We consider the most probable form of defects, i.e., the anionic vacancy in these materials. The corresponding diagram of $\Delta G_{H*}$ is plotted in Figure 6. For all these materials, their perfect sheets have an unfavorable HER activity reflected by the very large $\Delta G_{H*}$ exceeding 1.42 eV. Indeed, introduction of atomic vacancies promotes the adsorption of hydrogen for all the cases. However, the HER activity is still unfavorable due to the over binding of hydrogen at graphene ($\Delta G_{H*}$ of -1.61 eV) and very weak adsorption at atomic sites around the vacancies for these monochalcogenides with $\Delta G_{H*}$ more than 0.5 eV. Therefore, we find that the introduction of atomic vacancies does not give rise to satisfying HER activity due to either strong binding or weak adsorption of hydrogen (see more details in Figure S2 and Table S7 in Supporting Information). In contrast, the vacancy in phosphorene shows a very high HER activity with $\Delta G_{H*}$ of -0.02 eV (Table 1), complying with the thermoneutral condition ($\Delta G_{H*} \approx 0$). Therefore, we believe that our work provides a new opportunity for achieving high HER activity in 2D materials via defect creation and utilization. Also phosphorene, as a unique fundamental and exfoliatable 2D material with a direct band gap, could allow a direct utilization of solar energy for HER activity where atomic defects must play a critical role.

Concerning the edges, we only considered the armchair and zigzag edges so far. Recent experiments showed that other complex edges also existed in phosphorene. For example, it was shown that the edges of phosphorene adopted irregular eye-like shapes.[44] Another study revealed that the edges of phosphorene adopted a combination of different crystalline edges.[45] Herein, we performed additional calculations on these intermediate edge orientations between armchair and zigzag edges as shown in Figure 7. The edge orientation **c** that forms an angle θ with respect to zigzag direction can be defined through a pair of integers ($n_1$, $n_2$) with **c**=$n_1$**a**+$n_2$**b**, which is quite similar to the graphene edge. In particular, θ=0° (90°) is for zigzag (armchair) direction. The HER activities of those edges (both the pristine and reconstructed edges) are investigated by examining the adsorptions of hydrogen atoms and calculating their corresponding $\Delta G_{H*}$ (see Figure S3 and Table S8 in Supporting Information). It can be seen that for each type of edge, atomic sites with high HER activity also exist, suggesting that other types of phosphorene edges are also eligible for high HER activity in addition to armchair/zigzag edges.

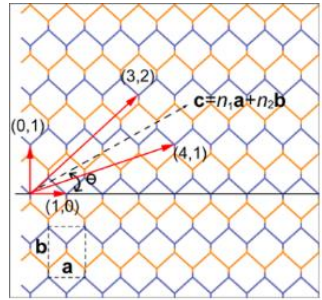

**Figure 7**. Construction of intermediate edges between zigzag and armchair directions of phosphorene. The edge orientation **c** that forms an angle θ with respect to zigzag direction is defined through a pair of integers ($n_1$, $n_2$) with **c**=$n_1$**a**+$n_2$**b**. In particular, θ=0° (90°) corresponds to zigzag (armchair) direction.

We would like to point out that bare phosphorene is known to suffer from structural degradation in air, which was previously ascribed to the effects of water and oxygen molecules. Recent experiments, however, proved the stability of phosphorene in deoxygenated water[21] and demonstrated the exfoliation of phosphorene using stabilizing surfactants in deoxygenated water.[22] Our simulation shows that the dissociation of water molecule at the vacancy site is unlikely owing to the high splitting barrier of 2.14 eV (see Figure S4 and Movie 1 in Supporting Information). For potential applications to HER, care has to be taken to avoid poisoning of phosphorene catalyst by $O_2$. Adding of an ionophore coating layer above phosphorene would be an effective approach to separate phosphorene from $O_2$, which was found to significantly increase the stability of phosphorene and allows only certain type of molecules/ionic species to selectively permeate through the coating layer.[46] Recently developed approaches like electron-coupled-proton buffer or proton exchange membrane,[47,48] which were adopted to avoid $H_2$/$O_2$ mixing during water splitting, may also be helpful to increase the stability of phosphorene by suppressing $O_2$ and by selectively passing through protons. Our findings here are not only useful for manipulating defects in phosphorene (vacancy and edge) to achieve a superior HER efficiency, but also potentially useful for exploring the HER activity for other phosphorus allotropes consisting of pentagon/hexagon rings, which are essentially the same as the core structures of reconstructed vacancies in phosphorene. Finally, our findings also highlight the importance of the activity of defects in *p*-orbital-dominated catalysts (such as recently reported layered $C_3N$ sheet[49] and borides/carbides[50]), which often possess a fast carrier kinetics and good reaction efficiency owing to their small effective mass associated with dispersive *p*-states and wide bands.

**Conclusion**

In summary, the non-metal character of phosphorene, together with its robust direct band gap, can realize solar-driven hydrogen production, allowing a direct transfer of solar energy to chemical energy. Recently, several experimental groups independently reported HER activity of phosphorene induced by infrared/visible light (Ref. 24 and Ref. 25) and a strong photocatalytic activity.[51] This triggers much interest in applying this material as photocatalysis for producing hydrogen renewably from solar energy. To our best knowledge, phosphorene is thus far the only 2D material that exhibits the direct conversion of solar to chemical energy. In this work, using density functional theory calculations, we provide a first attempt to uncover the catalytic mechanism from the atomic point of view. We show that the physical origin for the HER activity of phosphorene arises from defects: atomic vacancies and edges. These defects account for the superior HER activity through hydrogen compensation of these defects and modified H-P hybridization arising from the local lattice distortions of these defects. We report that similar atomic vacancies in some isostructural layered semiconducting materials are inactive, suggesting the unique functioning mechanism of the defects in phosphorene. Our findings are useful for guiding experiments for improving the HER performance of phosphorene via properly engineering these defects and shed light on the design of new *p*-orbital dominated photocatalysts for high HER performance.

## ASSOCIATED CONTENT

**Supporting Information**

Model building and computational details of first-principles calculations; details of Gibbs free energy calculation for HER; Implementation and parameters of kinetic Monte Carlo simulations; Coverage effects; Comparison of the phosphorene with TMDs and phosphorene-like materials with respect to the HER activity; HER at low-index intermediate edges between armchair and zigzag directions; MD simulation for the interaction of $H_2O$ and the MV at 300 K (Movie 1). This material is available free of charge via the Internet at http://pubs.acs.org.

## AUTHOR INFORMATION

**Corresponding Author**
zhangg@ihpc.a-star.edu.sg; zhangyw@ihpc.a-star.edu.sg
**Notes**
The authors declare no competing financial interests.

## ACKNOWLEDGMENT

This work was supported in part by a grant from the Science and Engineering Research Council (152-70-00017). The authors gratefully acknowledge the financial support from the Agency for Science, Technology and Research (A*STAR), Singapore and the use of computing resources at the National Supercomputing Centre Singapore (NSCC), Singapore.

TOC:

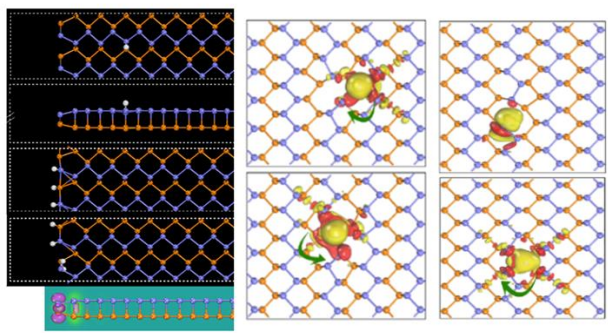